\documentclass[12pt]{iopart}

\usepackage{color}
\usepackage{graphicx}

\usepackage{enumerate}
\usepackage{comment}
\expandafter\let\csname equation*\endcsname\relax
\expandafter\let\csname endequation*\endcsname\relax
\usepackage{amsmath}
\usepackage{amssymb}
\usepackage[dvipsnames]{xcolor}
\usepackage{tikz}
\usepackage{amsthm}
\usepackage{pict2e}
\usepackage{bbold}
\usepackage{empheq}
\usepackage{subfigure}
\usepackage{xfrac}
\usepackage{hyperref}
\usepackage{enumitem}
\usepackage{notes2bib}
\usepackage[numbers]{natbib}
\usepackage{float}

\newcommand{\limacon}{Lima\c{c}on}

\begin{document}

\title[Steering internal and outgoing electron cavity dynamics]{Steering 
internal and outgoing electron dynamics in bilayer graphene cavities by cavity design}

\author{Lukas Seemann}
\address{Institute of Physics, Technische Universit\"at Chemnitz, D-09107 Chemnitz, Germany}
\ead{lukas.seemann@physik.tu-chemnitz.de}

\author{Angelika Knothe}
\address{Institut f\"ur Theoretische Physik, Universit\"at Regensburg, D-93040 Regensburg, Germany}

\author{Martina Hentschel}
\address{Institute of Physics, Technische Universit\"at Chemnitz, D-09107 Chemnitz, Germany}

\vspace{10pt}
\begin{indented}
\item[]March 4, 2024
\end{indented}

\begin{abstract}
Ballistic, gate-defined devices in two-dimensional materials offer a platform for electron optics phenomena influenced by the material's properties and gate control. We study the ray trajectory dynamics of all-electronic, gate-defined cavities in bilayer graphene to establish how distinct regimes of the internal and outgoing charge carrier dynamics can be tuned and optimized by the cavity shape, symmetry, and parameter choice, e.g., the band gap and the cavity orientation. In particular, we compare the dynamics of two cavity shapes, o'nigiri, and Lima\c{c}on cavities, which fall into different symmetry classes. We demonstrate that for stabilising regular, internal cavity modes, such as periodic and whispering gallery orbits, it is beneficial to match the cavity shape to the bilayer graphene Fermi line contour. Conversely, a cavity of a different symmetry than the material dispersion allows one to determine preferred emission directionalities in the emitted far-field. 

\end{abstract}

\section{Introduction}

Over the last years, electron optics in graphene-based systems has been an active field of research  \cite{chakraborti2024electron} due to the fascinating opportunities offered by the material's electronic structure, the high sample quality \cite{Bolotin2008, Du2008, Young2009, banszerusBallisticTransportExceeding2016, Rickhaus2013}, and the possibilities to control charge carriers via gating \cite{Novoselov2004, huard2007transport, lemme2007graphene, williams2007quantum}. Efforts have been undertaken, e.g., to investigate the transmission across graphene p/n junctions \cite{katsnelsonChiralTunnellingKlein2006, PhysRevB.85.085406, cheianovSelectiveTransmissionDirac2006, RickhausThesis2015},  magnetic focussing \cite{Taychatanapat2013, Bhandari2016, Chen2016}, electron guiding \cite{Rickhaus2015c, Rickhaus2015d, Graef2019}, collimation, and lensing \cite{cheianovFocusingElectronFlow2007, Liu2017, Lee2015, Barnard2017}, as well as the dynamics of graphene cavities, often termed "Dirac billiards" \cite{xuChaosDiracElectron2018, xuRelativisticQuantumChaos2021, hanDecaySemiclassicalMassless2018, huangPerspectivesRelativisticQuantum2020, Zhao2015, brunGrapheneWhisperitronicsTransducing2022, schneiderDensityStatesProbe2014, schneiderResonantScatteringGraphene2011, bardarsonElectrostaticConfinementElectrons2009, schrepferDiracFermionOptics2021}.

Recently, electron optics in bilayer graphene (BLG) has attracted particular attention. Unlike the Dirac cones of monolayer graphene, BLG's bands feature triangularly deformed bands due to trigonal warping already at low energies. This symmetry breaking translates into the real-space dynamics of electrons and holes, leading to three preferred propagation directions of the charge carriers per valley \cite{peterfalviIntrabandElectronFocusing2012, goldCoherentJettingGateDefined2021, inglaaynes2023ballistic}. Furthermore, in BLG, gating allows inducing a band gap at the K-points. Such gate-controlled band gap opening facilitates confinement of the charge carriers and entails generalised Fresnel laws for scattering at a potential step different from the Klein tunnelling at graphene p/n junctions \cite{milovanovicBilayerGrapheneHall2013, peterfalviIntrabandElectronFocusing2012, vanduppenFourbandTunnelingBilayer2013, maksym2023exact, SeemannKnHe_BLGI_PRB}. 

The Drosophila to study unusual ballistic charge carrier dynamics have been circular, all-electronic cavities in BLG, where the scattering region is defined by a gate-defined potential step \cite{schrepferDiracFermionOptics2021, SeemannKnHe_BLGI_PRB}. Already such a circular cavity, which is feasible to realise experimentally with current sample qualities \cite{banszerusBallisticTransportExceeding2016, leeBallisticMinibandConduction2016, goldCoherentJettingGateDefined2021, berdyuginMinibandsTwistedBilayer2020}, fabrication and gating techniques \cite{overwegElectrostaticallyInducedQuantum2018, overwegTopologicallyNontrivialValley2018, goldCoherentJettingGateDefined2021, iwakiriGateDefinedElectronInterferometer2022, rodan-legrainHighlyTunableJunctions2021, devriesGatedefinedJosephsonJunctions2021, portolesTunableMonolithicSQUID2022, dauberExploitingAharonovBohmOscillations2021, elahiDirectEvidenceKleinantiKlein2022, inglaaynes2023ballistic, Ingla-Aynes2023}, gives rise to a complex, mixed phase space for the ballistic trajectory dynamics featuring largely chaotic dynamics, regular islands of stable, triangular trajectories inside the cavity (induced solely by the $C_3$ symmetry of the trigonally warped BLG dispersion), and whispering gallery (WG)-type modes, which remain close to the cavity boundary for a certain number of scattering events before disappearing into the chaotic sea \cite{SeemannKnHe_BLGI_PRB}.

Here, we go beyond these previous works on circular electronic BLG cavities and investigate how cavity dynamics is influenced by cavity symmetry and design for cavity shapes other than circular ones. To this end, we chose two exemplary cavity shapes, the o'nigiri cavity (which closely resembles the $C_3$ symmetric Fermi lines in BLG) and the \limacon{} cavity (which is mirror symmetric and does not respect $C_3$ symmetry). For both cavity shapes, we analyse the internal dynamics of trajectories inside the cavity and the far-field generated by trajectories emitted from the cavity.

For the \emph{internal cavity dynamics}, we find that the $C_3$ symmetric o'nigiri cavity helps to promote some of the features already present for circular cavities: For example, depending on the cavity orientation with respect to the BLG lattice (and, hence, to the Fermi line in momentum space), the o'nigiri cavity selectively stabilises or de-stabilises specific triangular orbits. In particular, one may choose a shape and orientation of the o'nigiri cavity, which maximises the lifetime of triangular and WG-type modes with a certain sense of orientation per valley. Conversely, a \limacon{} cavity destroys most regular motion inside the cavity. We hence find that for versatile control and tailoring of particular stable trajectories inside the cavity, it is beneficial to match the symmetry and shape of the cavity and the Fermi line. 

While the \emph{far-field} is mainly determined by the BLG dispersion outside of the cavity, we find that depending on the cavity shape and geometry, these outgoing states may be populated with different weights,  thereby skewing the emission pattern: The o'nigiri cavity allows to adjust the height of different emission peaks while preserving $C_3$ symmetry. Emission from a \limacon{} cavity, in turn, is no longer $C_3$ symmetric, with one of the three peaks per valley being more pronounced and defining a preferred emission direction.

This paper is organized as follows. In Sec.~\ref{sec_model}, we introduce the model of an all-electronic, gate-defined BLG cavity. We then study the internal trajectory dynamics in the o'nigiri cavity, focusing on how to employ the interplay of material and geometric properties to confine electron dynamics, e.g., to WG-like modes with long electron lifetime in the cavity. We compare to the internal dynamics in a \limacon{} cavity. We then turn to the dynamics of electrons that have left the cavity in Sec.~\ref{sec_extdyn}. First, we relate the far-field emission characteristics to the cavity's phase space properties of the internal dynamics and continue by analyzing the effects of geometry and bandgap for the o'nigiri cavity. In Sec.~\ref{sec_lima}, we analyze the additional far-field control provided by an axially symmetric \limacon{} cavity. We close with a Summary in Sec.~\ref{sec_summary}.

\begin{figure}[h]
    \centering
    \includegraphics[scale=1]{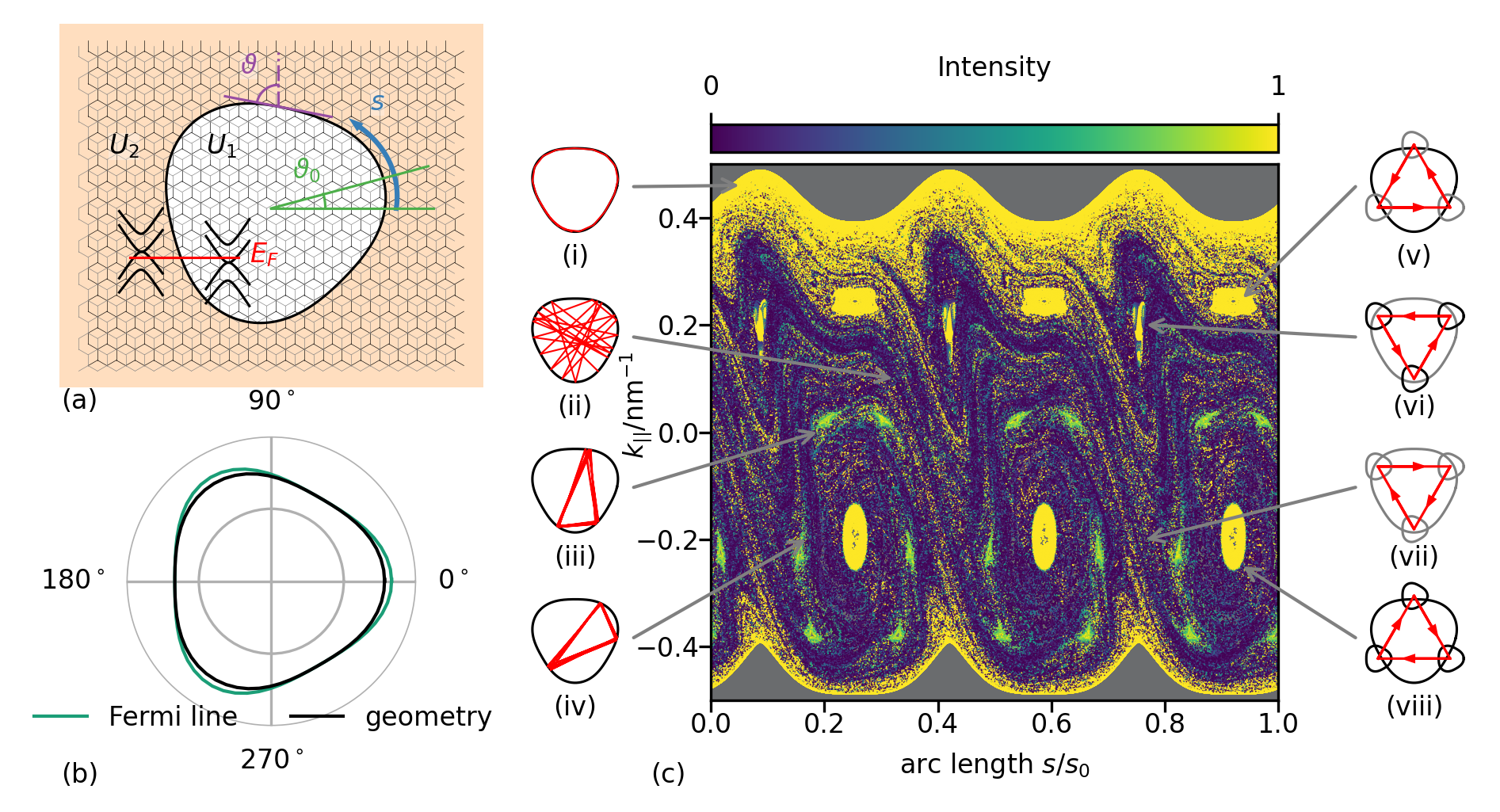}
    \caption{(a) Schematic of the model system. The electronic cavity (black outline) is defined by the potential step $U_0$ between $U_1$ (white region) and $U_2$ (orange region) corresponding to a shift in the energy bands (inset), $E_F$ is the Fermi energy. The orientation angle $\vartheta_0$ corresponds to the tilt angle between the symmetry axis of the cavity and the BLG lattice. The angle $\vartheta$ defines the orientation of the boundary tangent relative to the lattice. The arc length $s$ is measured from the x-axis along the cavity boundary. Note that the ratio of lattice and cavity sizes is not to scale. (b) Comparison of the Fermi line and o'nigiri cavity geometry with  $\varepsilon=0.08$ ($\vartheta_0 = 0^\circ$),showing a similar curvature at the flat sides that crucially determine the non-chaotic electron dynamics. (c) PSOS (central panel) of the electron dynamics in an o'nigiri shaped BLG cavity with $\varepsilon_3=0.08$, $\vartheta_0=30^\circ$, $E_F=160\,$meV, $U_0=170\,$meV and $\Delta=0\,$meV. For this and the other PSOS, we started $10000$ trajectories and traced their scattering points and reflected intensity for (up to) $100$ reflections. The small panels (i)--(iv) depict characteristic trajectories and their position in PSOS. Panels (v)--(viii) illustrate  the interplay of real-space (cavity geometry) and momentum-space (Fermi line contour) based (in)stabilities, indicated by black (gray) lines. In (viii), a coincidence of stabilities implies an increase of island size in the PSOS, whereas coinciding instabilities lead to unstable fixed points in (vii). The combination of stable and unstable situations, as in (v, vi), can either destabilize orbits (shrunk islands) that are stable in the circular cavity (vi) or create stable islands for originally unstable electron orbits (vi).  
}
    \label{fig:model_system} \label{fig:PSOS_onigiri_stab_instab}
\end{figure}
 
\section{Tuning the internal electron dynamics in bilayer graphene cavities}
\label{sec_model}

We consider gate-defined, all-electronic ballistic BLG cavities of various shapes that we characterise based on the charge carrier trajectory dynamics. The electronic properties of BLG subject to a gate-induced potential landscape can be described in terms of the four-band Hamiltonian,  \cite{mccannLandauLevelDegeneracyQuantum2006, mccannLowEnergyElectronic2007, mccannElectronicPropertiesBilayer2013, SeemannKnHe_BLGI_PRB},

\begin{equation}
 H^{\xi}_{BLG}=\xi
 \left(
 \begin{array}{cccc}
  \xi U-\frac{1}{2}\Delta & v_3\pi & v_4 \pi^{\dagger} & v \pi^{\dagger}\\
  v_3 \pi^{\dagger}&\xi U+\frac{1}{2}\Delta & v\pi & v_4 \pi\\
  v_4 \pi & v\pi^{\dagger} & \xi U+\frac{1}{2}\Delta  & \xi \gamma_1\\
  v\pi & v_4 \pi^{\dagger} & \xi \gamma_1 &\xi U-\frac{1}{2}\Delta 
 \end{array}
 \right),
\label{eqn:H}
\end{equation}

  in the sublattice basis $(A,B^{\prime},A^{\prime}, B)$ in the valley $K^{+}$ ($\xi=1$) and $(B^{\prime},A, B,A^{\prime})$ in the valley $K^{-}$ ($\xi=-1$)  
with $\pi=p_x+i p_y$, $\pi^{\dagger}=p_x-i p_y$. Here, $\Delta$ is an interlayer asymmetry gap, and $U(\mathbf{r})=U_0\mathcal{H}(\mathbf{r})$ is the potential step of height $U_0$ locally defining the cavity boundary (where $\mathcal{H}$ denotes the Heaviside step function).  Further, we include the couplings $v\approx 1.02 \cdot 10^6$ m/s, $v_3 \approx 0.12\cdot v$, $v_4\approx 0.37 \cdot v_3$, and $\gamma_1\approx 0.38$ eV \cite{kuzmenkoDeterminationGatetunableBand2009}. 

Notably, the skew hopping parameter between the layers, $v_3$, breaks rotational symmetry and induces trigonal warping in the Fermi line. This anisotropy in momentum space induces preferred propagation directions of the charge carriers at any given energy in real space \cite{goldCoherentJettingGateDefined2021, chakraborti2024electron, peterfalviIntrabandElectronFocusing2012, inglaaynes2023ballistic}. In Ref.~\cite{SeemannKnHe_BLGI_PRB}, we discussed how such collimation of the charge carriers' velocities leads to unusual trajectory dynamics in circular electronic BLG cavities. In particular, we found stable triangular cavity modes related to the preferred propagation directions induced by the BLG dispersion. Further, the breaking of rotational symmetry implies that WG-like trajectories in the vicinity of the cavity boundary, a characteristic and paradigmatic class for isotropic systems (with circular Fermi line and most optical systems), cannot be stable even if the cavity geometry is perfectly circular. In the following, we address the changes in the charge carrier dynamics when deforming the BLG cavity.
 
The all-electronic gate-defined BLG cavity we consider is schematically shown in Fig.~\ref{fig:model_system}(a). The cavity geometry is defined by a potential step from $U_1$ (inside) to $U_2$ that confines the electronic charge carriers. We parametrize the cavity orientation by a global orientation angle $\vartheta_0$ (between the lattice and the cavity symmetry axis) and a local boundary tilt angle $\vartheta$ (describing the local orientation of the cavity boundary with respect to the lattice). To calculate the anisotropic Fresnel laws for scattering at this boundary, we use this tilt angle to transform into a coordinate system with its $y$ axis aligned parallel to the cavity boundary tangent for each scattering event of an electron at the cavity boundary and solve the corresponding wave matching problem \cite{SeemannKnHe_BLGI_PRB, katsnelsonChiralTunnellingKlein2006, tudorovskiyChiralTunnelingSinglelayer2012, snymanBallisticTransmissionGraphene2007, nilssonTransmissionBiasedGraphene2007,  nakanishiRoleEvanescentWave2011, milovanovicBilayerGrapheneHall2013, parkBandGapTunedOscillatory2014, nakanishiTransmissionBoundaryMonolayer2010, barbierKronigPenneyModelBilayer2010, vanduppenFourbandTunnelingBilayer2013, sanudoStatisticalMagnitudesKlein2014, saleyKleinTunnelingTriple2022, heChiralTunnelingTwisted2013, maksym2023exact}. The arc length describes the position of each scattering position along the cavity boundary. In the following, we use the normalized arc length $\Tilde{s}=s/s_0$, where $s_0$ is the circumference of the cavity.

{\it Phase-space representation of asymmetric BLG cavities. }
To discuss the additional effects of a non-circular BLG cavity on top of the previously discussed material anisotropies, we chose different cavity geometry deformations with distinct symmetries. In polar coordinates, we parametrize the cavity geometry as

\begin{equation}
    R(\varphi) = R_0(1+ 
    \varepsilon_n\cos \, n(\varphi -\vartheta_0))\:,
    \label{eq:geo_shape}
\end{equation}

where $R_0$ is the mean radius of the cavity and $\varepsilon_n$ is a deformation parameter. We consider two different geometries given by  $n=1$ and $n=3$. The $n=3$ case we refer to as the o'nigiri geometry derived from the so-called shortegg shape \cite{shortegg}. The o'nigiri shape is threefold rotationally symmetric and resembles the $C_3$ symmetric trigonally warped BLG Fermi line, cf.~Fig.~\ref{fig:model_system}(b). This resemblance will be of special interest when addressing the interplay of real and momentum space orbit-(de)stabilizing effects, see Fig.~\ref{fig:model_system}(c) and its discussion below. 
For $n=1$, we obtain the mirror-symmetric so-called \limacon{}   shape \cite{limacon} that has attracted significant interest as it allows to realize microcavity lasers \cite{limacon_Susumu_Taka2009,limacon_Kim2009,limacon_Cao2009,limacon_NJPhys,limacon_Albert2012}.

For these different cavity symmetries, we explore the internal electron cavity dynamics in terms of trajectories that we represent in the so-called Poincaré surface of section (PSOS). The PSOS comprises, for each scattering event at the cavity boundary, the normalized arc length $\tilde{s}$ and the momentum $k_\parallel$ parallel to the boundary that is a conserved quantity at each scattering event. We perform this analysis of the electron trajectory dynamics using a generalised ray-tracing algorithm developed in Ref.~\cite{SeemannKnHe_BLGI_PRB} that takes into account the material-specific, anisotropic velocity distribution and Fresnel laws. Notice that due to these generalised, anisotropic Fresnel-laws \cite{SeemannKnHe_BLGI_PRB}, the angles of incidence and reflection are not equal.  

For the o'nigiri geometry, we illustrate the interplay between cavity parameters (shape and orientation) and Fermi line morphology and its impact on the internal electron dynamics in Fig.~\ref{fig:PSOS_onigiri_stab_instab}(c)\footnote{In the following, we will mainly focus on the $\xi=1$ valley and will discuss the effect of the other valley only where relevant. We provide the corresponding data for the other valley, $\xi=-1$, in Appendix \ref{sec:appendix}.}
The main panel shows the PSOS  for a non-circular o'nigiri geometry with $\varepsilon_3=0.08$ and $\vartheta_0 = 30^\circ$. We observe a phase space structure that consists of so-called islands (roundish objects) implying regular dynamics (triangular orbits corresponding to the charge carriers' three preferred propagation directions \cite{SeemannKnHe_BLGI_PRB}) and scattered dots indicating a chaotic motion. The prominent yellow belt on the upper and the somewhat smaller belt at the lower boundary of the PSOS correspond to counterclockwise and clockwise traveling WG-type trajectories that orbit close to the cavity boundary before they eventually escape into the chaotic region. The color scale indicates the intensity of electron trajectories remaining inside the cavity according to the reflection coefficients that can be derived from Eq.~(\ref{eqn:H}) via wave matching \cite{SeemannKnHe_BLGI_PRB}. 

The PSOS in Fig.~\ref{fig:PSOS_onigiri_stab_instab}(c) is not symmetric with respect to the $k_{\parallel}=0$ line, a characteristic difference to the circular geometry \cite{SeemannKnHe_BLGI_PRB}. While for a circular cavity, all islands and structures would be of the same size, for the deformed cavity of Fig.~\ref{fig:PSOS_onigiri_stab_instab}, some of the islands related to the triangular orbits increase in size, whereas others shrink. We also observe the appearance of additional islands not present in circular cavities.

The phase space structure in Fig.~\ref{fig:PSOS_onigiri_stab_instab}(c) can be understood as follows. The Fermi line's threefold ($C_3$) symmetry leads to stable triangular orbits even for circular cavity shapes. Their propagation directions are the three preferred group velocities corresponding to regions of small curvature of the trigonally warped Fermi line. The stability of these triangular orbits is hence a \textit{momentum-space based stability}.
In addition and somewhat similarly, the o'nigiri cavity geometry provides areas of small curvature in real space, which induce regular dynamics on triangular orbits and stable islands in the PSOS. These features induced by the cavity geometry hence correspond to \textit{real-space based stability}. Depending on how we combine these kinds of stability and the instabilities originating from the areas of high curvature in the Fermi line and cavity geometry, respectively, we can enlarge or shrink the stable islands in the PSOS and even change the character of unstable islands to stable. 

We demonstrate the interplay of the Fermi lines, the corresponding trajectories, and the cavity shape in the right panels (v)-(viii) in Fig.~\ref{fig:PSOS_onigiri_stab_instab}(c). These sketches illustrate the triangular orbit (red line) related to the structure indicated by the arrow (black). The shape enclosing this orbit is the real space (o'nigiri) geometry. This cavity shape is shown in black if its orientation stabilizes the triangular orbit; otherwise, it is shown in grey (real-space-based stability). The smaller forms at each reflection point indicate the Fermi line. Its orientation is fixed in space (as is the BLG lattice orientation). The Fermi line is shown in black when its preferred propagation directions agree with those of the triangular orbit and grey otherwise (momentum-space-based stability).   

We observe that the islands of the stable triangular orbit (viii) are prominent for $k_\parallel < 0$. Here, the reflection points of the stable triangular orbit lay on the flat sides of the o'nigiri geometry, and the stabilization effects of the momentum and the real space add up. For $k_\parallel > 0$, the stable triangular orbit (viii) is destabilized by reflecting at regions with high curvature of the cavity, yielding significantly smaller islands. In contrast, the flat sides of the o'nigiri cavity stabilize the unstable triangular orbit (vi) (that has no corresponding island structure in circular cavities), which results in a chain of three islands around the unstable fixed points of the circular cavity. Notice that this additional chain of islands does not appear for $k_\parallel < 0$ for the orbit (vii) that remains unstable after adding up two destabilizing effects.

It is a paradigm of nonlinear dynamics that the stable (islands) and unstable fixed points organize the overall phase space structure \cite{lichtenberglieberman}. We hence discuss the effect of the triangular-orbit island chains and their varying size on the other trajectories, such as WG-type orbits. Indeed, the chain of six islands for $k_\parallel > 0$ represents a barrier \cite{partialbarriers} to the motion of trajectories in phase space and provides relatively efficient confinement of the WG-type trajectories with $k_\parallel > 0 $ (counterclockwise movement). The effect of this boundary can be seen in Fig.~\ref{fig:PSOS_onigiri_stab_instab}(c) by the pronounced band of high intensity for the highest $k_\parallel$ values. In contrast, the large islands for $k_\parallel < 0$ do not represent an efficient barrier for clockwise propagating WG type orbits because of the existence of unstable fixed points in between the islands related to expanding directions in phase space and thus not supporting WG-like trajectories. Hence, an appropriate choice of parameters can enlarge the area of high-intensity WG-like trajectories.

\label{sec_internal}

\begin{figure}[h]
    \centering
    \includegraphics{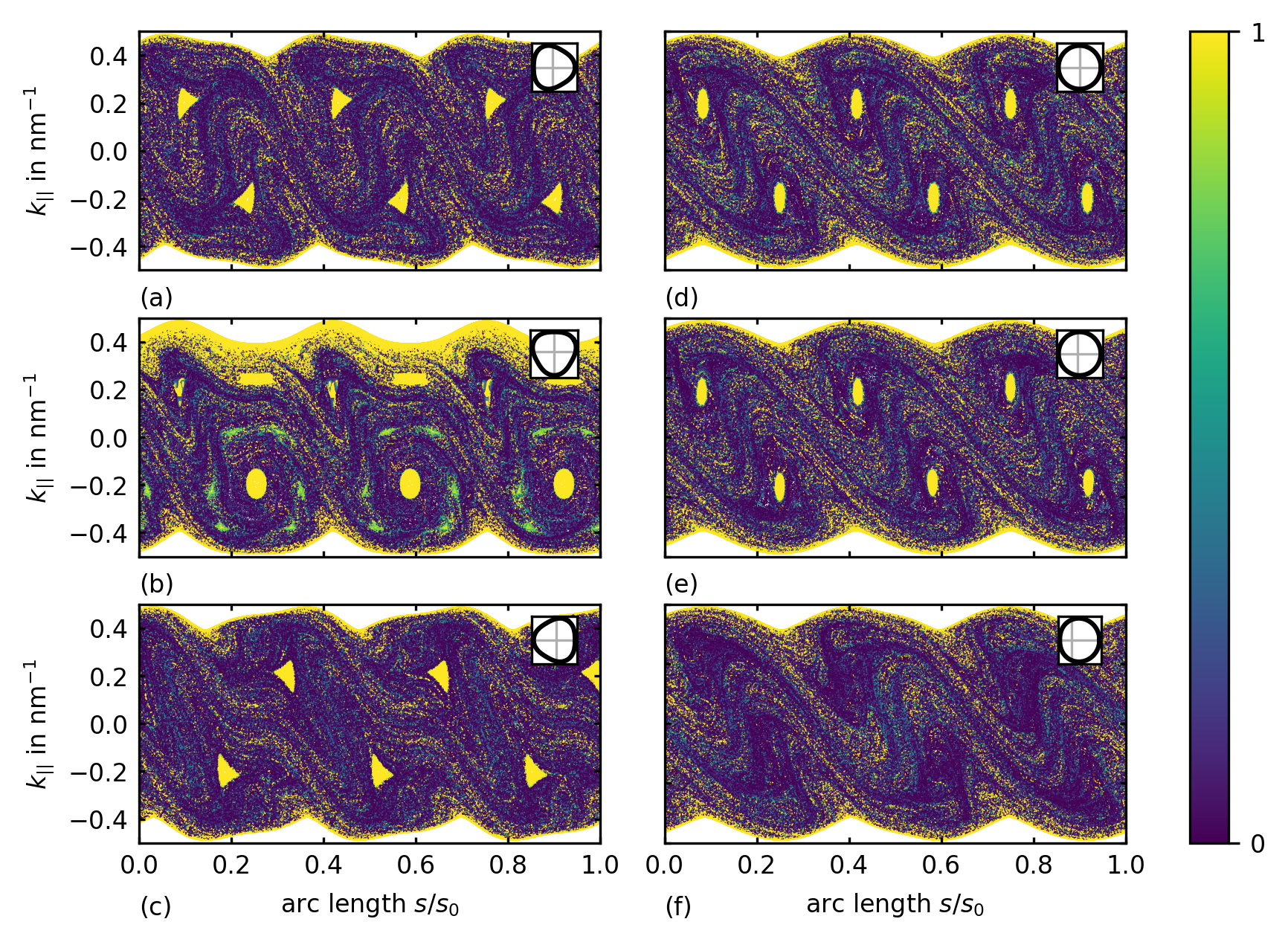}
    \caption{PSOS for different geometries and orientation angles $\vartheta_0$, respectively. The insets show the shape and the orientation of the cavity. Shown are  o'nigiri cavities with geometry parameter $\varepsilon_3=0.08$ and orientation angle (a) $\vartheta_0=0^\circ$, (b) $\vartheta_0=30^\circ$ and (c) $\vartheta_0=60^\circ$ as well as (d) a circular cavity and (e,f) Lima\c{c}on-shaped cavities with (e) $\varepsilon_1=0.1$ (slight deviation from the circle) and (f) $\varepsilon_1=0.4$. The color scale represents the intensity of rays remaining inside the cavity.}
    \label{fig:pr_overview}
\end{figure}
 
{\it Geometry dependence of the PSOS.}
In Fig.~\ref{fig:pr_overview}, we illustrate different PSOS for various cavity shapes and orientation angles $\vartheta_0$.  
Figure \ref{fig:pr_overview}(a,b,c) show the PSOS of o'nigiri shaped cavities with $\varepsilon_3=0.08$ and $\vartheta_0=0^\circ,30^\circ,60^\circ$. Note that Fig.~\ref{fig:pr_overview}(b) is the case discussed above. The PSOS depends strongly on the orientation angle $\vartheta_0$ as it affects the conditions for the momentum and real space stabilities to enforce or counteract each other. The islands of the stable triangular orbit, discussed in \cite{SeemannKnHe_BLGI_PRB} for circular cavities and shown in Fig.~\ref{fig:pr_overview}(d), are distinctly visible. 
Figure \ref{fig:pr_overview}(e,f) illustrates the PSOS of two different Lima\c{c}on shaped  cavities. We observe that the islands can be destroyed for a large deformation parameter (here $\varepsilon_1=0.4$ in (f)) and that the mismatch of the (momentum space) symmetry of the Fermi line and the (real space) symmetry of the cavity geometry yields a further source of chaotic dynamics: when the symmetry of the Fermi line does not match that of the geometry, there is no possibility to stabilize orbits as described above for the o'nigiri cavity; rather, the interplay of the two non-commensurate symmetries favors chaotic (irregular) trajectories.

{\it Optimizing phase space for whispering gallery type orbits.}
In optical cavities where light is confined by total internal reflection, the life time or survival probalitiy of WG-like  modes is a figure of merit.
Similarly, the survival properties of charge carriers in WG-type modes are of special interest in the context of electronic transport and applications since they are directly related to the life time of charge carriers in the cavity. Here, we analyze how to choose an optimal cavity shape in order to maximize or minimize the degree of WG-like orbits and their lifetime inside the cavity.

\begin{figure}[h]
    \centering
    \includegraphics{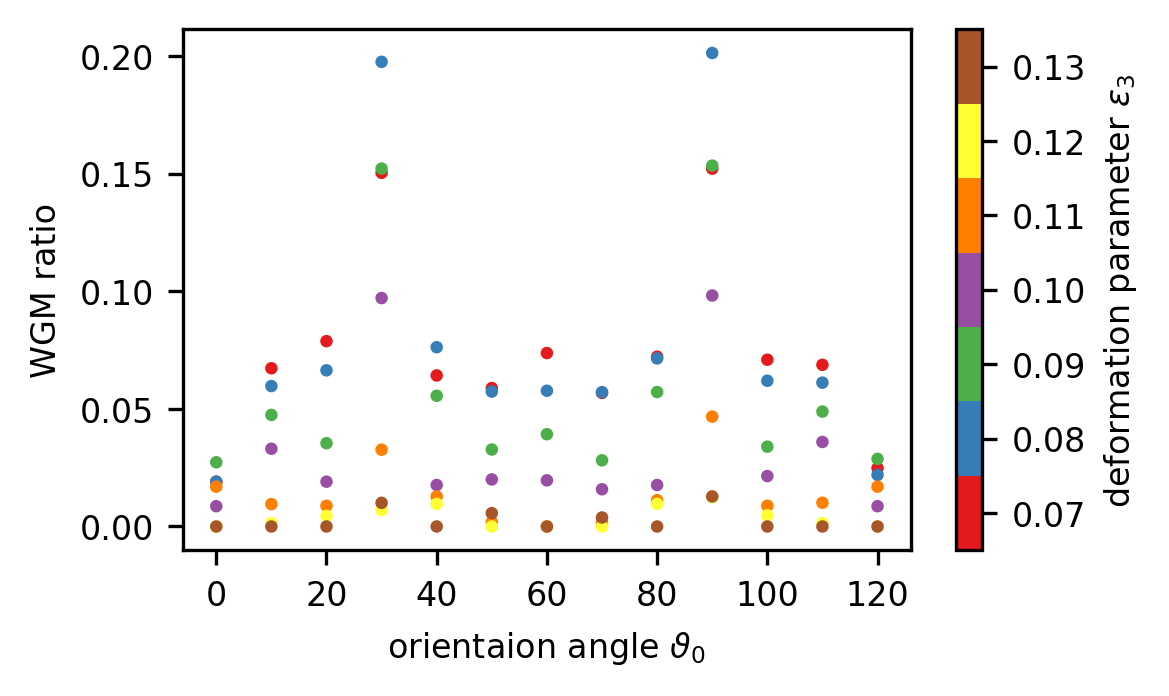}
    \caption{The relative number of WG-like trajectories in an o'nigiri-shaped cavity depending on the orientation angle $\vartheta_0$ for different deformation parameters $\varepsilon_3$ (the color scale corresponds to the value of the deformation parameter). There are distinct maxima for $\vartheta=30^\circ$ and $\vartheta=90^\circ$. We identify $\varepsilon_3=0.08$   as the optimum parameter based on the color scale. We point out the similarity of the (flat side) curvatures of the Fermi contour and the cavity geometry for this value of $\varepsilon_3$, cf.~Fig.~\ref{fig:model_system} (c).}
    \label{fig:wgm_opti}
\end{figure}

In order to quantify the lifetime of WG-like modes, we start 
$300$ trajectories with random initial conditions and unit intensity and trace them for $100$ reflections. We call a trajectory a WG-like trajectory if (i) its intensity after $100$ reflections remains larger than 0.95 and (ii) its  WG degree, which is the ratio of the trajectory segment between the subsequent scattering point and the corresponding arc length along the cavity boundary, is always higher than 0.8. After $100$ reflections, we determine the WGM ratio, indicating the number of WG-like trajectories over the total number of trajectories. 
In Fig.~\ref{fig:wgm_opti}, the WGM ratio is plotted over the deformation parameter $\varepsilon_3$ and the orientation angle $\vartheta_0$. The various points for each $\vartheta_0$ in Fig.~\ref{fig:wgm_opti} correspond to different deformation parameters $\varepsilon_3$. The positions of the maxima determine the optimal choice of the parameters to ensure a large portion of WG-like trajectories. We find that a deformation parameter of $\varepsilon_3 \approx 0.08$ and orientation angles $\vartheta_0$ of 30$^\circ$ or 90$^\circ$ are the optimal values that favor the presence of WG-type trajectories. While the optimal $\vartheta_0$ follows from symmetry arguments, the shape parameter $\varepsilon_3$ takes its optimal value when the curvature of the Fermi line and the cavity geometry match along the flat sides, cf.~Fig.~\ref{fig:model_system}(b). Hence, we find that matching the cavity shape and the Fermi line is optimal for the efficient confinement of WG-type modes. 

These optimal parameter values are close to the parameters chosen for Fig.~\ref{fig:PSOS_onigiri_stab_instab}, where the stabilization of counterclockwise traveling WG-type orbits is indeed visible due to an increase in the number of stable islands protecting the WG-type trajectories. Here, the main contribution stems from stabilizing the unstable triangular orbit of the o'nigiri cavity geometry by the Fermi-line induced preferred propagation directions that coincide with those of that trajectory.
 
\section{Electron far-field emission of the o'nigiri cavity}
\label{sec_extdyn}

In the following, we will focus on the dynamics of electrons emitted from the BLG cavity. We will elucidate how the internal dynamics translates into the external electron dynamics and investigate the effects of various system parameters, e.g., the (convex or concave) contour of the Fermi line,  the shape ($\varepsilon_{1,3}$) and orientation ($\vartheta_0$) of the geometry as well as the band gap ($\Delta$) which jointly determine the emission characteristics of the BLG cavity.

\subsection{Convex vs.~concave Fermi line} 

We first analyse the influence of the Fermi line contour because the preferred propagation directions induced by a trigonally warped Fermi line rule the external dynamics as well as the internal dynamics. For the external dynamics, the shape of the  Fermi line in the BLG region outside the cavity is crucial.  
For the trigonally warped Fermi lines of BLG, two cases must be distinguished: a concave and a convex Fermi line \cite{SeemannKnHe_BLGI_PRB}, cf.~Fig.~\ref{fig:model_system}(b) and Fig.~\ref{fig:in_out_transl}(c) that illustrate the convex and concave Fermi line obtained for different values of the potentials $U_1$ (inner Fermi line) and $U_2$ (outer Fermi line), respectively.
A convex outer Fermi line results in three preferred emission directions corresponding to the flat sides of the Fermi line, while a concave outer Fermi line leads to a bifurcation of each of the three peaks. We will discuss examples for both cases in detail below (cf.~Figs.~\ref{fig:in_out_transl}(b,c)  entailing Fig.~\ref{fig:PSOS_FF_Limacon_040}) for bifurcated and non-bifurcated far fields). 
Indeed, it is the shape of the Fermi line that mainly determines the structure of the far field. However, various other effects and controllable parameters can be employed to change the (relative) intensity of a single far-field peak or the ratio between the double peaks. 
 
\subsection{Cavity geometry: deriving far-fields from the cavity's phase-space portrait}

The influence of the cavity geometry on the internal dynamics was discussed in Sec.~\ref{sec_internal}. Here, we will study how these properties translate into the far-field emission characteristics.  
To this end, we consider an o'nigiri-shaped cavity as in Fig.~\ref{fig:PSOS_onigiri_stab_instab} with $\varepsilon=0.08$ and orientation angle $\vartheta_0=30^\circ$. Since the o'nigiri provides the same $C_{3}$ symmetry as the Fermi line, also the far-field will be $C_{3}$ symmetric. For a convex Fermi line, this three-fold rotational symmetry will simply lead to three emission peaks, similar to the emission pattern from a circular cavity \cite{SeemannKnHe_BLGI_PRB}. Here, we focus on the more interesting case of a concave Fermi line and demonstrate that it is possible to adjust the ratio between the double peaks via the cavity orientation $\vartheta_0$. 

\begin{figure}[htb]
    \centering
    \includegraphics{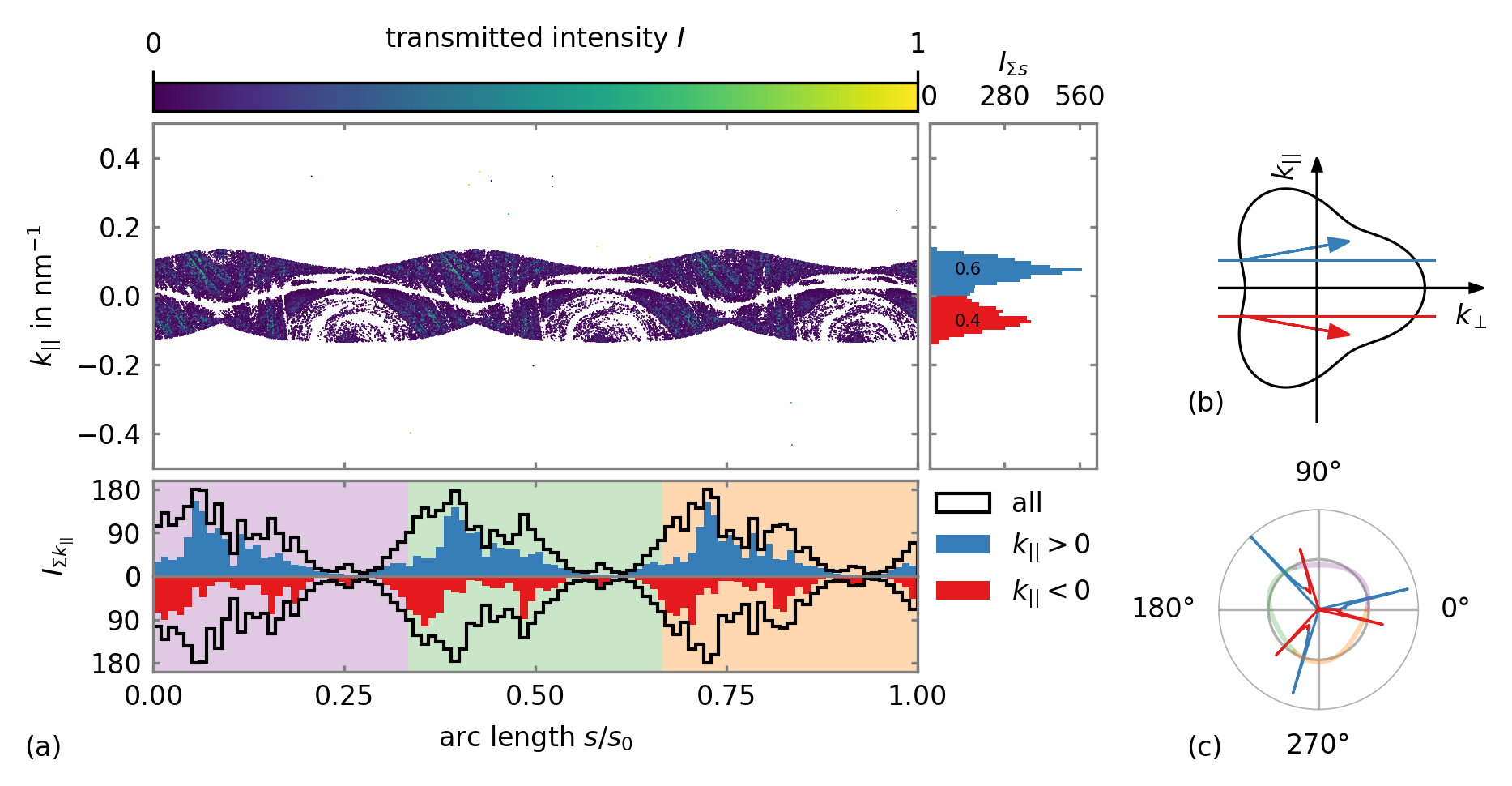}
    \caption{Relation between PSOS and far-field emission. In (a), we show the PSOS of Fig. \ref{fig:PSOS_onigiri_stab_instab}  but with the transmitted (rather than the reflected) intensity as a color scale featuring a rich structure that determines the external dynamics. We bin the summed transmitted intensity with regard to the arc length $s/s_0$ and the $k_{||}$ component of the wave vector, respectively. The $k_{||}$ component of the wave vector translates into the group velocities as the normal vector to the Fermi line (b). The horizontal lines in (b) correspond to the two peak maxima in the histogram. Their different heights result in various prominent bifurcation features in the far field (c). The cavity's shape and orientation are shown in pastel colors in the far field plot (c), corresponding to the highlighted areas in the arc length histogram.}
    \label{fig:in_out_transl}
\end{figure}

\begin{figure}[htb]
    \centering
    \includegraphics{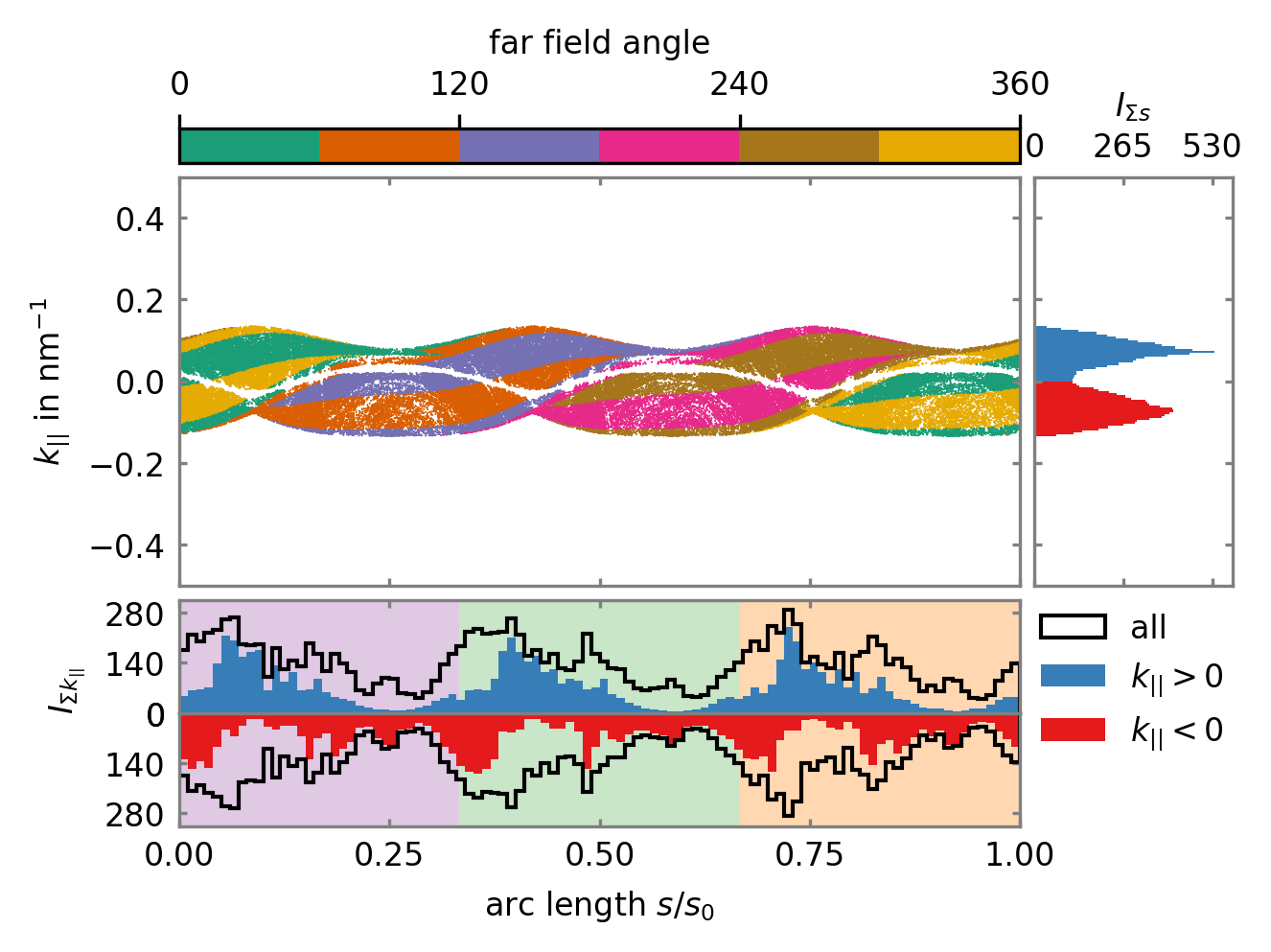}
    \caption{Relationship between the PSOS and the far-field emission angles (same parameters as in Fig. \ref{fig:in_out_transl}, $\varepsilon_3=0.08$ and $\vartheta=30^\circ$). We plot the far-field angle of the transmitted electrons as a color scale. The different colors can be used to identify the origin of the emitted electrons in phase space. The color changes at $0^\circ$, $120^\circ$ and $240^\circ$ correspond to the bifurcation of the far-field peaks.}
    \label{fig:PSOS_chi_ff}
\end{figure}

Figure \ref{fig:in_out_transl} illustrates how the internal cavity dynamics translates into the emitted far field. The transmitted intensity for each scattering event is plotted as a color scale in the PSOS of Fig.~\ref{fig:PSOS_onigiri_stab_instab}. By summing up and binning the transmitted intensity with regard to the arc length $s/s_0$ and the $k_{||}$ component of the wave vector, respectively, the main directions of transmission can be identified. The asymmetry in the $k_{||}$ histogram directly determines the asymmetry of the bifurcated peaks in the far field. The extensive islands ruling the internal dynamics for $k_{||}<0$ in Fig.~\ref{fig:PSOS_onigiri_stab_instab} capture substantial intensity on stable orbits inside the cavity. Consequently, the charge carriers with $k_{||} > 0$ contribute more to the transmission, e.g.~via WG-like trajectories that eventually leave the cavity. Since the far-field bifurcation peaks correspond to different signs of $k_{||}$, see Fig.~\ref{fig:in_out_transl}(b), one of the two peaks is favored whenever there exists a $\pm k_{||}$ asymmetry, cf.~Fig.~\ref{fig:in_out_transl}(c). 

We further investigate the connection between the internal and outgoing dynamics by studying in which far-field direction (angle) the electron is emitted at each boundary scattering event, cf.~Fig.~\ref{fig:PSOS_chi_ff}. To this end, we color-code the far-field angle distinguishing the six distinct emission regions for better visibility.  
The emission angles form a characteristic pattern in phase space, with certain areas of the PSOS contributing to certain emission directions. Such a pattern is similar to the optical case \cite{limacon}, where the unstable manifold (or steady probability distribution) provides a measure related to the chaotic cavity dynamics (of light rays) and rules the far-field emission characteristics. In the case of BLG cavities, we additionally find a strong impact of the non-circular Fermi line on the far-field emission properties. For the concave Fermi line considered here, we point out that the bifurcation of the far-field peaks can be directly traced back to the color changes at $0^\circ$ (yellow to green), (orange to purple), and $240^\circ$ (pink to brown).

\begin{figure}[h]
    \centering
    \includegraphics{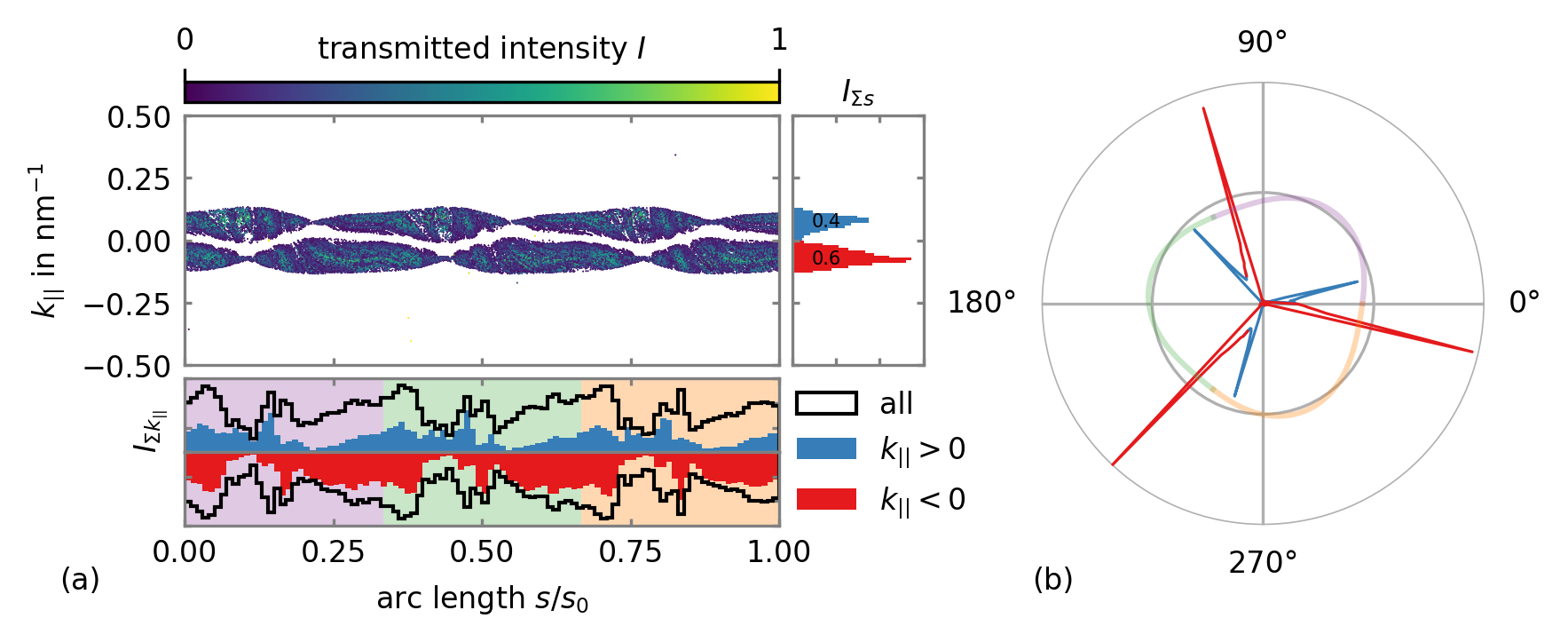}\\
    \includegraphics{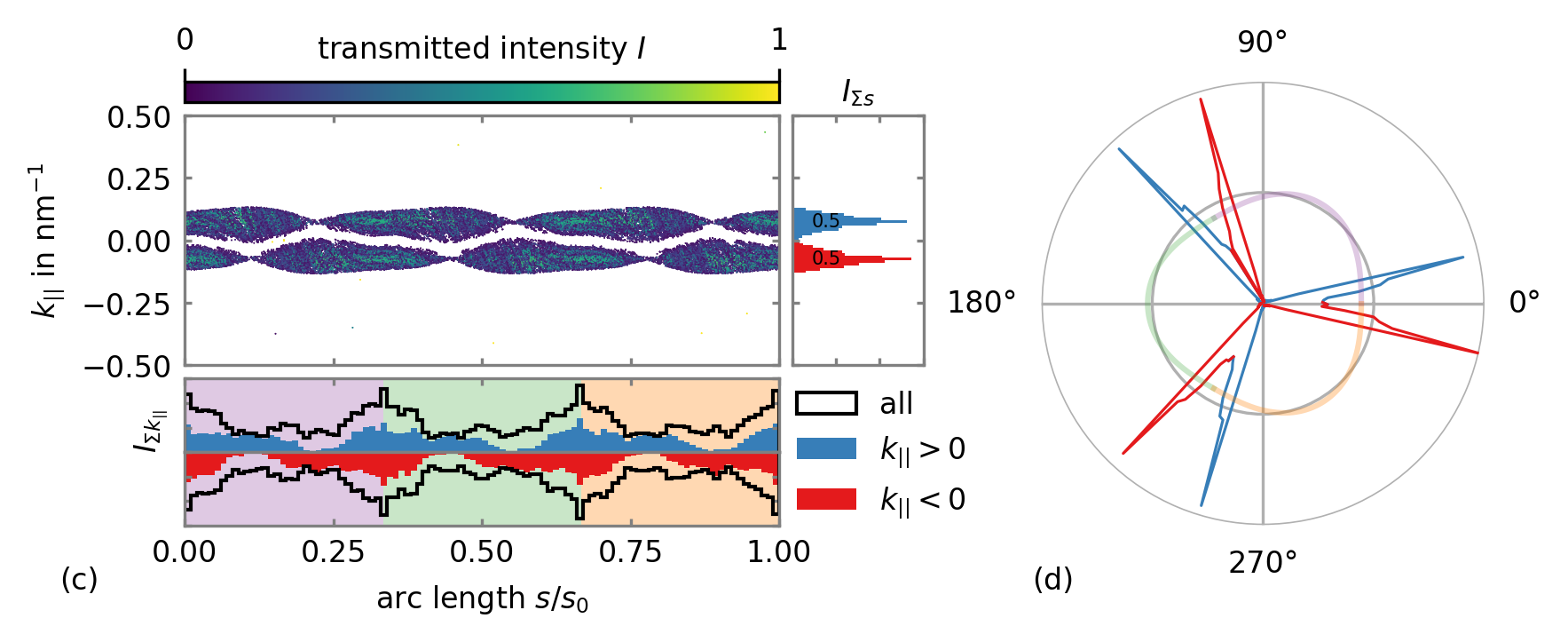}\\
    \includegraphics{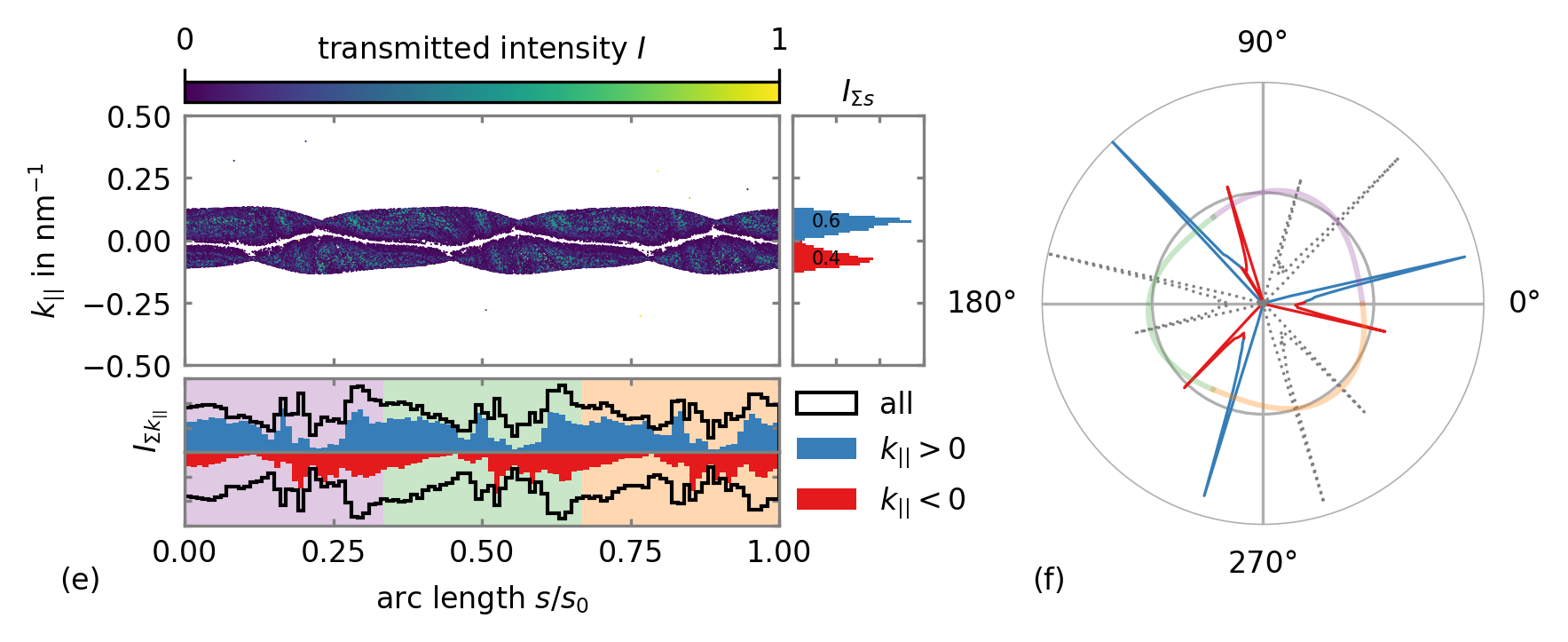}
    \caption{Influence of the orientation angle $\vartheta_0$ on the far-field emission properties of o'nigiri shaped BLG cavities ($\Delta=0\,\mathrm{meV}$, $\varepsilon_3=0.08$ and (a,b) $\vartheta_0=50^\circ$, (c,d) $\vartheta_0=60^\circ$, (e,f) $\vartheta_0=70^\circ$. (a,c,e) Transmitted intensity weighted PSOS and histograms for transmitted electrons (color scale) and (b,d,f) far-field emission. We indicate the cavity shape in (b,d,f) in pastel colors matching those in the lower panels of (a,c,e). The relative peak height within the three emission double peaks is controlled by $\vartheta_0$. The dashed line in panel (f) indicates the far-field emission resulting from the $K^-$ valley. The slight differences in the peak height are of numerical origin.}
   \label{fig:PSOS_onigiri_eps080_th50_70_delta0}
\end{figure}
 
For the non-circular symmetric cavities, the orientation angle $\vartheta_0$ between the cavity and the BLG lattice further influences the dynamics. In Fig.~\ref{fig:PSOS_onigiri_eps080_th50_70_delta0}, we provide the far fields for three different orientation angles $\vartheta_0$. In each case, the $k_{||}$ histogram of the transmitted intensity indicates the different weighting of the bifurcation peaks. We show the most symmetric far field with equal bifurcation peaks in Fig.~\ref{fig:PSOS_onigiri_eps080_th50_70_delta0}(d) for $\vartheta_0=60^\circ$. By varying the orientation angle $\vartheta_0$, it is possible to suppress each of the double peaks selectively, cf.~Fig.~\ref{fig:PSOS_onigiri_eps080_th50_70_delta0}(b) and (f). 

In Fig.~\ref{fig:PSOS_onigiri_eps080_th50_70_delta0}(f), the far-field emission of the other ($\xi=-1$) valley is indicated as a grey dashed line. The cavity's overall emission thus consists of several emission directions that can be more or less pronounced. Due to the possible sequence of two weaker followed by two more substantial peaks, the mean emission directions might appear, depending on the double peak splitting and the resolution, to be rotated, e.g. by 30$^\circ$ in Fig.~\ref{fig:PSOS_onigiri_eps080_th50_70_delta0}(f). 

\subsection{Influence of the band gap $\Delta$}

\begin{figure}[htb]
    \centering
    \includegraphics{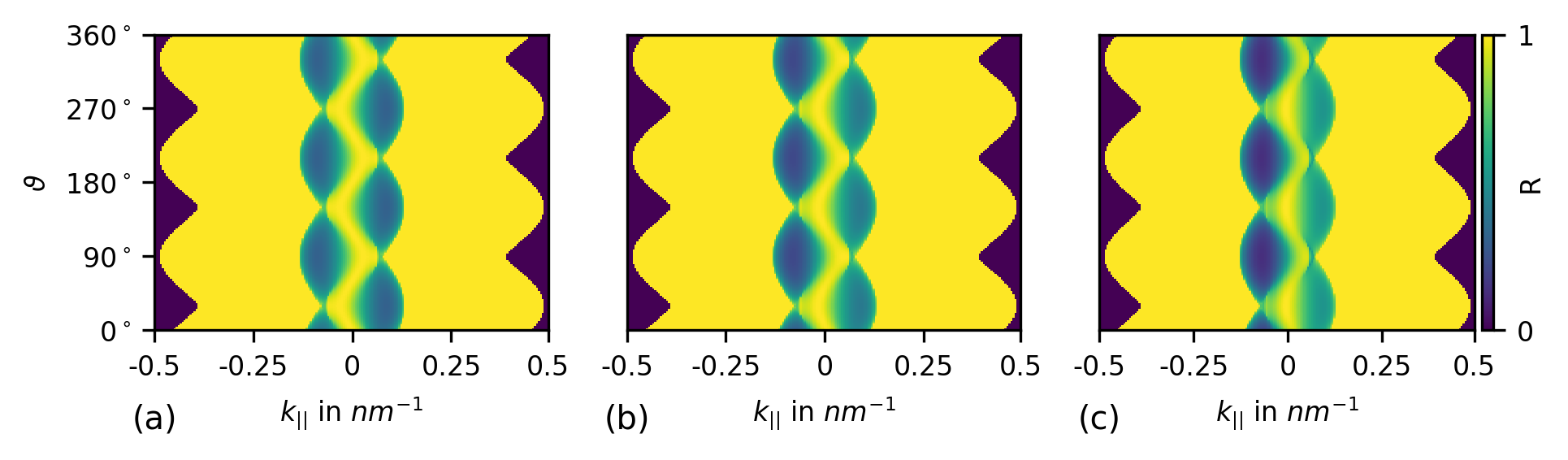}
    \caption{Reflectivity coefficient $R$ in the $k_\parallel$ 
-$\vartheta$ plane for different band gaps, $\Delta$, (a) $\Delta=0\,\mathrm{meV}$, (b) $\Delta=5\,\mathrm{meV}$, and (c) $\Delta=10\,\mathrm{meV}$, for a circular cavity with $E_F=160\,$meV and $U_0=170\,$meV. A finite band gap breaks the $\pm k_\parallel$ symmetry of the reflectivity and induces an intravalley asymmetry already for a circular cavity.
    }   
   \label{fig:map_0-5-10} 
\end{figure}

The relative height of the double peaks can further be adjusted by tuning the band gap $\Delta$ for any given cavity geometry. We illustrate the underlying mechanism in Fig.~\ref{fig:map_0-5-10}, where the reflection coefficient $R$  obtained from the wave matching calculations is given in the $k_\parallel - \vartheta$ plane for various $\Delta$. Whereas $R$ is fully symmetric with respect to $k_\parallel =0$ for vanishing bandgap $\Delta=0$, cf.~Fig.~\ref{fig:map_0-5-10}(a), a band gap induces an asymmetry between positive and negative $k_\parallel$, cf.~Fig.~\ref{fig:map_0-5-10}(b,c). As the sign of $k_\parallel$ is related to clockwise and counterclockwise propagation direction within the cavity, we observe an asymmetry between right and left moving modes per valley for finite $\Delta$. In the other valley, the role of right and left moving trajectories is reversed. Therefore, trajectory-reversal symmetry is restored once trajectories in the other valley are included.
 
The origin of this band-gap-dependent imbalance of the reflection coefficient is a broken symmetry in the Hamiltonian. Without band gap, $\Delta=0$, the Hamiltonian Eq.~(\ref{eqn:H}) is invariant under the transformation $k_y \rightarrow -k_y$ in connection with a change of the basis $(A,B',A',B)\rightarrow(B',A,B,A')$. 
This change is feasible due to the inter-layer sublattice equivalence, which, however, is broken when a band gap $\Delta>0$ is applied. Consequently, the reflectivity $R$ in each valley separately is not $\pm k_y$ (or $\pm k_\parallel$) symmetric \cite{FBandTunBLG2013, BLGElecBarriers2009}. 

\begin{figure}[htb]
    \centering
    \includegraphics{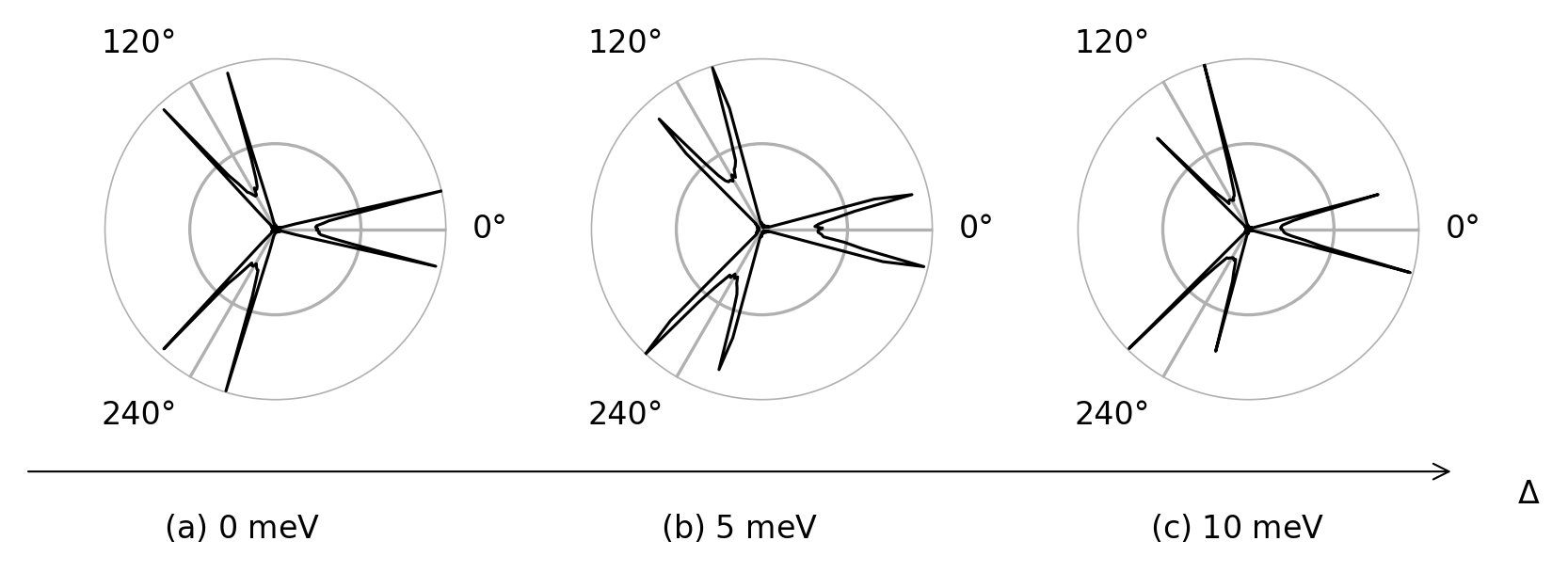} \\
    \vspace{0.5cm}
    \includegraphics{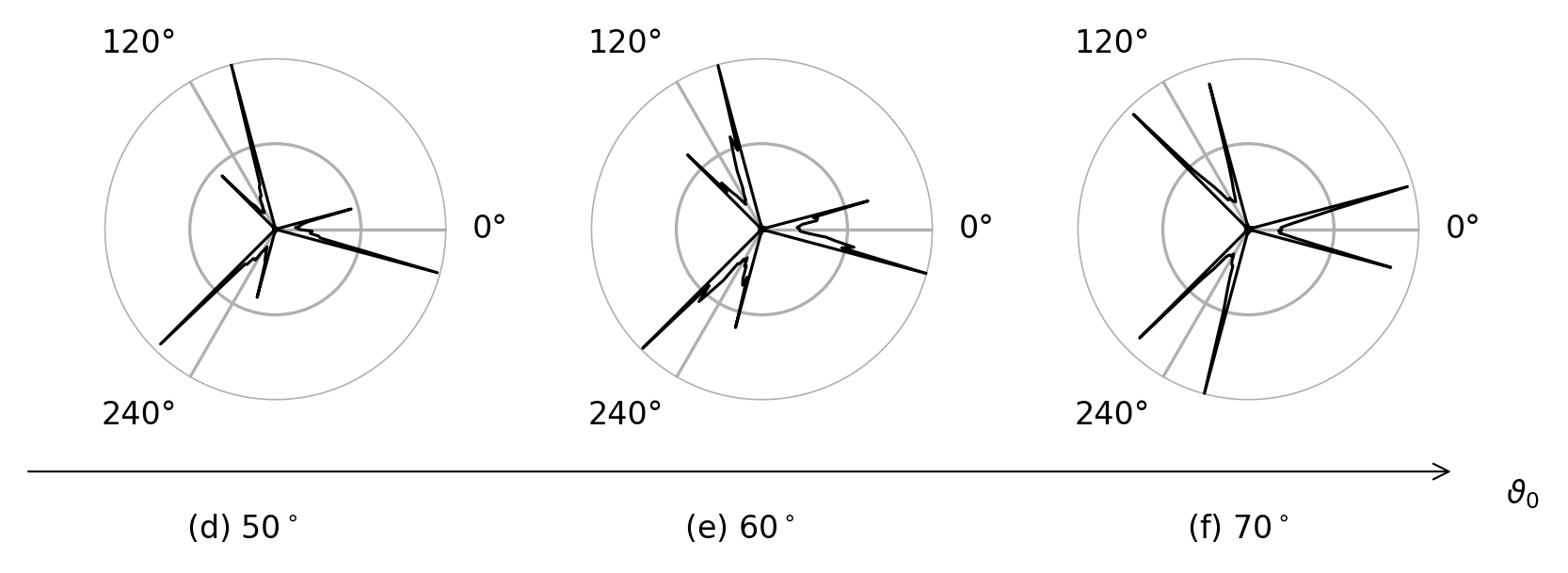}
    \caption{Far-field emission characteristics for different band gaps $\Delta$ and various cavity geometries. (a)-(c) Circular cavities with (a) $\Delta = 0$ meV, (b) $\Delta = 5$ meV, and (c) $\Delta = 10$ meV showing an increasing asymmetry in the emission pattern. (d)-(f) Far fields of o'nigiri cavities with $\Delta=10\,\mathrm{meV}$ and orientation angles (d) $\vartheta_0=50^\circ$, (e) $\vartheta_0=60^\circ$, and (f) $\vartheta_0=70^\circ$. The geometry-induced imbalance of the double peaks (cf.~Fig.~\ref{fig:PSOS_onigiri_eps080_th50_70_delta0}), can be increased (d) or reversed (f) by superposing band-gap induced asymmetries in $R$.}
    \label{fig:doublepeakfiniteDelta}
\end{figure}

In the case of a concave Fermi line, each of the double peaks corresponds to a positive or negative $k_{\parallel}$ value, respectively. Consequently, without a band gap, $\Delta=0$, the peaks are equivalent.
However, a finite band gap $\Delta > 0$ favors transmissions with negative $k_{||}$ values since the reflectivity $R$ is lower for $k_{||} < 0$ (see Fig.~\ref{fig:map_0-5-10}). Therefore, an asymmetry in the double peaks arises when inducing a finite band gap. Figure \ref{fig:doublepeakfiniteDelta}(a-c) illustrates how one of the double peaks can be suppressed in a circular cavity by applying a finite band gap $\Delta$.

Combining the effects of cavity deformations and the band gap discussed above, a finite $\Delta$ can be used to manipulate the cavity geometry effects on the far field. In Fig.~\ref{fig:doublepeakfiniteDelta}(d-f), we show the far fields of the o'nigiriy cavity for the same orientation angles as in Fig.~\ref{fig:PSOS_onigiri_eps080_th50_70_delta0}, but for a finite band gap  $\Delta=10\;$meV. The band gap-induced asymmetry in the transmittivity favors negative $k_{||}$ for transmission, hence suppressing the $k_{||}<0$ peak in the far field for finite $\Delta$.
This suppression can moreover be influenced by varying the cavity orientation with respect to the lattice, $\vartheta_{0}$ as is illustrated 
in Fig.~\ref{fig:doublepeakfiniteDelta} (d-f). 
 
Note the difference in the $\vartheta_0=60^\circ$ far fields for $\Delta=0$ in Fig.~\ref{fig:PSOS_onigiri_eps080_th50_70_delta0}(d) and for $\Delta=10$ meV in Fig.~\ref{fig:doublepeakfiniteDelta}(e).
We also point out the balanced peaks for $\vartheta_0=70^\circ$, cf.~Fig.~\ref{fig:doublepeakfiniteDelta}(f). For this choice of $\vartheta_{0}$, the band gap induced asymmetry favors the $k_{||}<0$ peak while the shape geometry  favors the $k_{||}>0$ peak. If the strength of both effects is equal, the symmetry between the peaks in the far field is restored (compare, e.g., the  $\Delta=0$-symmetric case, $\vartheta_0=60^\circ$, in Fig.~\ref{fig:PSOS_onigiri_eps080_th50_70_delta0}(d) to Fig.~\ref{fig:doublepeakfiniteDelta}(e)).
 
\section{Electron far-field emission of the Lima\c{c}on cavity}
\label{sec_lima}
We complement our considerations by studying the electron emission from a non-rotationally symmetric cavity. The cavity being of a different symmetry allows one to engineer far-field patterns that do not respect the $C_{3}$ symmetry of the BLG Fermi line. Specifically, we discuss the Lima\c{c}on geometry, Eq.~(\ref{eq:geo_shape}) with $n=1$, and use a convex Fermi line, such that we avoid peak splitting and may focus on the overall peak heights. The \limacon{} cavity is a well-known model system of mesoscopic optics because of its pronounced directional emission for higher deformation parameters around $\varepsilon_1=0.4$ \cite{limacon,limacon_Albert2012, limacon_Cao2009, limacon_Kim2009, limacon_NJPhys, limacon_Susumu_Taka2009}. Remarkably, this universal and, in optics, robust property is overtaken in BLG by the three-peak far-field structure induced by the trigonal Fermi line that also determines the three main emission directions. However, the geometric shape effect is sufficient to distort the magnitude of these three peaks. 
 
\begin{figure}
    \centering
     \includegraphics[height=0.25\textheight]{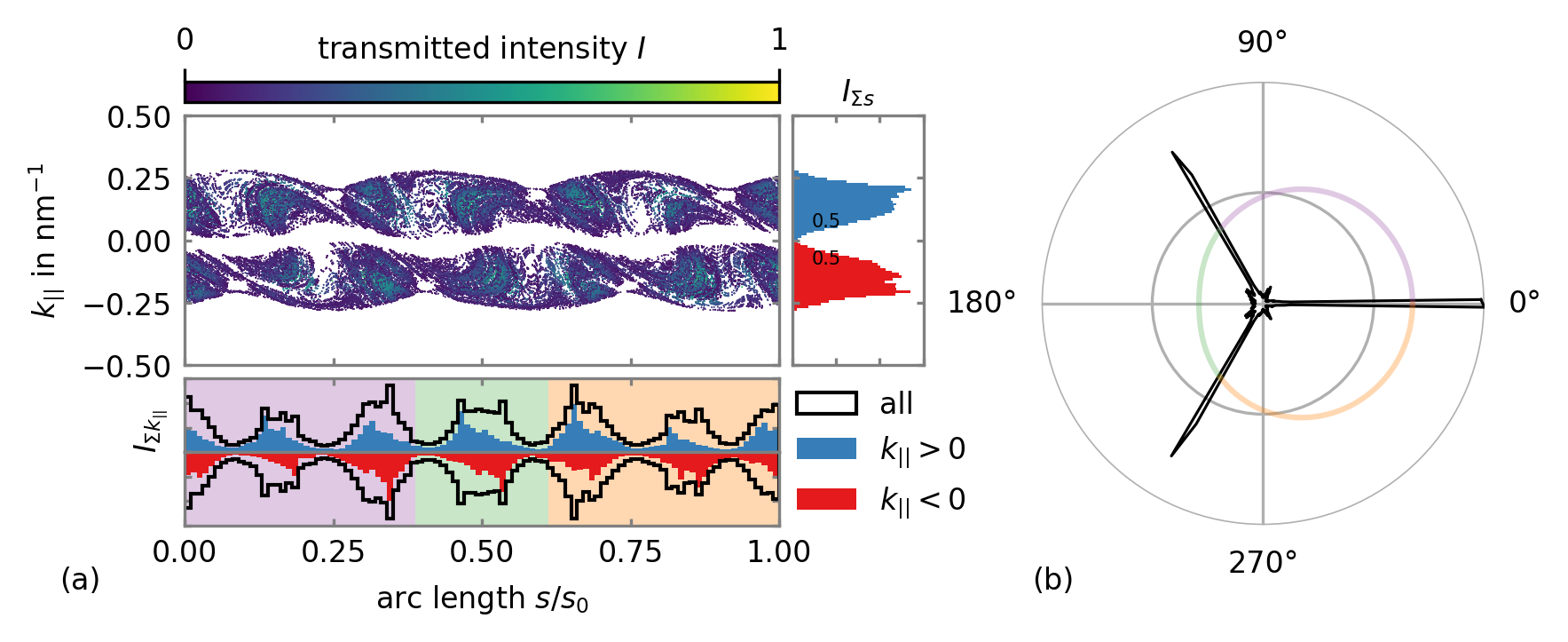}\\
     \includegraphics[height=0.25\textheight]{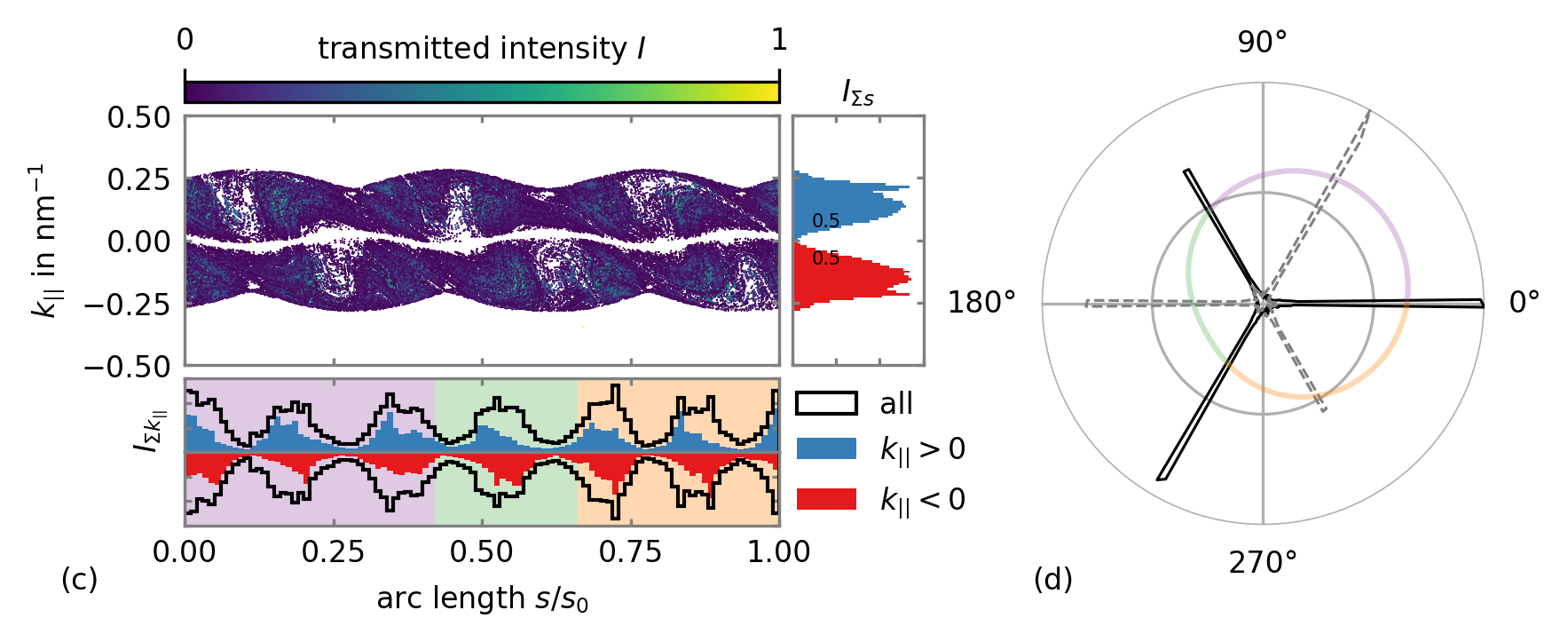}
    \caption{Far-field of the Lima\c{c}on cavity ($\varepsilon_1=0.4$) for two different orientation angles, (a,b)  $\vartheta_0 = 0^\circ$ and (c,d) $\vartheta_0 = 30^\circ$,  and a convex Fermi line ($E_F = 100\,$meV and $U_0 = 160\,$meV), so there is no bifurcation in the three main emission peaks (balanced $k_\parallel$ histogram). While the far field reflects the cavity's mirror symmetry for $\vartheta_0=0^\circ$ in (b), the symmetry is broken for $\vartheta_0=30^\circ$ in (d), leading here to the suppression of the far-field emission at $120^\circ$.} The dashed line illustrates the far-field distribution of the $K^-$ valley.
    \label{fig:PSOS_FF_Limacon_040}
\end{figure}

In our specific example, combining the mirror symmetry provided by the Lima\c{c}on shape and the $C_{3}$ symmetry of the Fermi line leads to three differently pronounced far field peaks of which their relative intensity depends on the shape and orientation of the Lima\c{c}on. We illustrate this far-field pattern in  Fig.~\ref{fig:PSOS_FF_Limacon_040} where we show the transmitted intensity and the far-field emission of Lima\c{c}on shaped cavities with $\varepsilon_1=0.4$ for two different orientation angles $\vartheta_0$ ($E_F=100\,\mathrm{meV}$ and $U=160\,\mathrm{meV}$, corresponding to a convex Fermi line). 
Depending on $\vartheta_0$,  different peaks in the far field are enlarged or suppressed. Selectively addressing individual peaks, inducing thus anisotropic far-field emission, is not possible for an o'nigiri-shaped cavity where both the cavity and the Fermi line respect $C_{3}$ symmetry.
The far-field emission for the other ($\xi=-1$) valley is given by the dashed line in Fig.~\ref{fig:PSOS_FF_Limacon_040}(d), illustrating the resulting complex total emission pattern. 

As before, we illustrate the origin of this effect using intensity-weighted histograms derived from the phase-space resolved transmitted intensity. 
Notice that the Lima\c{c}on geometries implies a distorted relation between polar angle and boundary arc length, resulting in non-equal arc length sections (pastel colors in Fig.~\ref{fig:PSOS_FF_Limacon_040}) for each of the three (Fermi-line induced) emission directions. Depending on the height of the histogram bar at the position of the color change, the corresponding far-field peak increases or decreases.

We note that for the convex Fermi lines considered in this section, considering a band gap $\Delta$  in the BLG spectrum does not lead to any recognizable deviations in the far-field emission for a Lima\c{c}on shaped cavity due to the chaotic nature of the internal electron dynamics that effectively averages out asymmetries in the reflectivity.

\section{Summary}
\label{sec_summary}
Using a trajectory analysis based on ray-wave correspondence, we have investigated the interplay and mutual influence of real and momentum space symmetry breaking on the internal and external (far field) electron dynamics in BLG cavities of various geometric shapes. We observe rich charge carrier dynamics and identify a versatile toolbox of different tuning parameters - ranging from applied gate voltages (determining the Fermi line contour and the band gap $\Delta$) via cavity shape tuning (inducing stable trajectories and favoring specific emission directions) to controlling the cavity symmetry axis with respect to the underlying BLG lattice (tilt or orientation angle $\vartheta_0$). We summarize our findings as follows.

The {\it internal} electron dynamics is ruled by the interplay between momentum-space based stable trajectories (here, the triangular orbits induced for a trigonally warped Fermi line, which are present even for circular cavities) and real-space based stable trajectories that depend on the cavity shape. For the o'nigiri cavity geometry, where the real and momentum space shapes are very similar, we demonstrate that such a \textit{shape similarity is particularly useful for tuning the system's dynamical properties}. The interplay between real and momentum space features can be used to change the size of stable islands or turn unstable fixed points into finite-size stable islands, thus structuring the PSOS in the desired way. Such tuning of the trajectory dynamics can be used, e.g., to stabilize WG-like orbits and ensure a long lifetime of charge carriers inside the cavity. Being able to affect the charge carrier dynamics inside a cavity can help tailor its transport properties, and a change in the external parameters can trigger the capture or release of electrons from the cavity.   Conversely, if the shape and symmetry of the cavity do not comply with those of the Fermi line, we find all regular internal dynamics to be destroyed, as we demonstrate for the example of Lima\c{c}on-shaped cavities. Hence, we conclude that to stabilize specific internal, regular trajectories, matching the Fermi line and the cavity shape is beneficial. 
 
The {\it external} charge carrier dynamics originates from electrons leaving the BLG cavity with an intensity according to (generalized Fresnel-type) transmission coefficients. The far-field carrier dynamics then depends on both the BLG dispersion and the cavity shape. 
The characteristics of the material imply the following features of the far-field emission:

(i) Independently of the cavity geometry, the trigonally warped BLG Fermi line in the region outside the cavity induces collimated, preferred emission directions.  While a convex BLG generally leads to three peaks in the emission, these peaks are further split if the Fermi line is a concave triangle. These narrow, collimated jets in the emission represent the main feature of the BLG cavity far field.  The band gap and the cavity shape and orientation can further tune the relative intensity of the peaks.

(ii) A band gap $\Delta$ induces an asymmetry between clockwise and counterclockwise propagating trajectories per valley. Such an asymmetry implies that the peaks within the double peak of one valley attain different weights. 
For a cavity shape other than circular, the combination of cavity symmetry and momentum space $C_{3}$ symmetry further affects the far-field emission:

(iii) The cavity geometry can be further used (besides the band gap $\Delta$) to tune the relative weight of the two peaks within a far-field emission double peak for concave Fermi lines.

(iv) Cavities that do not respect the symmetry of the BLG Fermi line can show emission spectra with other than $C_{3}$ symmetry. In particular,  the tilt angle $\vartheta_0$ between the cavity axis and BLG lattice allows the tuning of the relative height of the far-field peaks and the definition of the main emission directions of electrons leaving the BLG cavity. 

We close with a brief discussion of the possible implications of our results for future experimental studies of gate-defined BLG cavities. Ballistic BLG devices on the micrometer scale with different gate geometries are feasible with current experimental techniques \cite{banszerusBallisticTransportExceeding2016, leeBallisticMinibandConduction2016, goldCoherentJettingGateDefined2021, berdyuginMinibandsTwistedBilayer2020, overwegElectrostaticallyInducedQuantum2018,  iwakiriGateDefinedElectronInterferometer2022, portolesTunableMonolithicSQUID2022,   inglaaynes2023ballistic}. For a given shape of the gate-defined cavity, changing the Fermi energy by the gates allows adjusting the shape of the triangular Fermi lines, e.g., from concave to convex. We anticipate that stable, regular trajectories inside an electrostatic BLG cavity will lead to anisotropic transport through such cavities due to the bunching of trajectories in specific positions along the cavity boundary. Also, the distinct spatial distribution of different cavity modes (e.g., triangular trajectories vs.~WG-like modes) should make it possible to discern such modes, e.g., by their different coupling to contacts at different arc length positions around the cavity. Likewise, the anisotropic emission of the far field patterns may be detected, e.g., in transport or in scanning gate experiments. We hence propose all-electronic gate-defined BLG cavities as versatile platform to study ballistic electron optics phenomena where transport and coupling properties can be tuned and optimised.

\section{Acknowledgement}
We thank Síle Nic Chormaic from OIST, Okinawa, Japan, for pointing out the relation to the shape of the Japanese rice snack called o'nigiri. We acknowledge discussions with Samuel M\"oller, Christoph Stampfer, Leon Stecher, Carolin Gold, Klaus Richter, Sibylle Gemming, Síle Nic Chormaic, Thomas Busch, and Satoshi Sunada. AK acknowledges support from the  Deutsche Forschungsgemeinschaft (DFG, German Research Foundation) within  Project-ID 314695032 -- SFB 1277 and DFG Individual grant KN 1383/4. MH thanks the TU Chemnitz for support within the Visiting Scholar Program (Sìle Nic Chormaic). 

\clearpage

\section{Appendix}\label{sec:appendix}

In the main part of this study, we mostly focussed on one valley. Here we briefly add the results for the other valley, $\xi = -1$. 

\begin{figure}[H]
    \centering
    \includegraphics[width=0.6\textwidth]{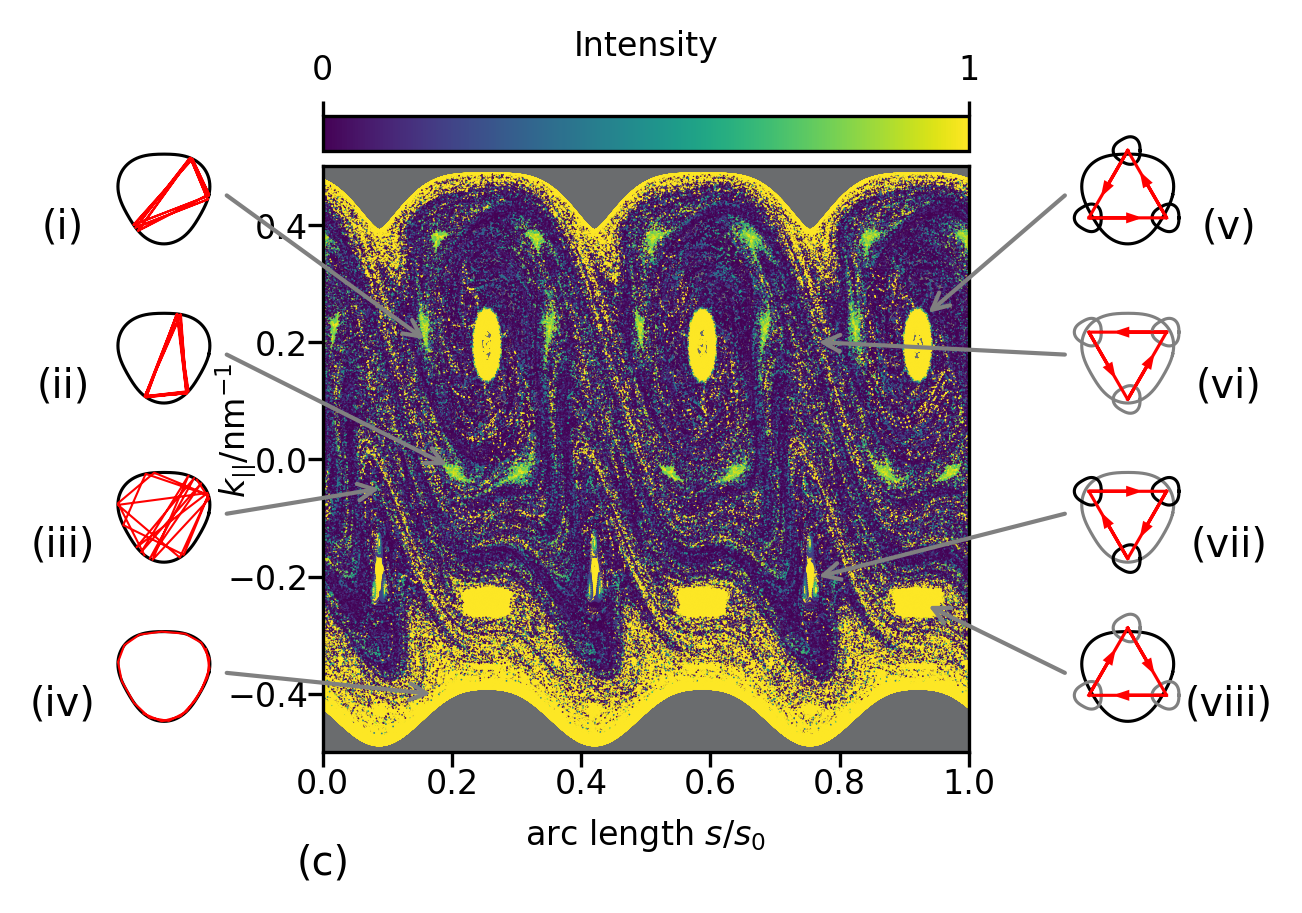}
    \caption{As Fig.~\ref{fig:model_system})(c), but for the other valley, $\xi=-1$. }
    \label{fig:app1}
\end{figure}

\begin{figure}[H]
    \centering
    \includegraphics[width=0.6\textwidth]{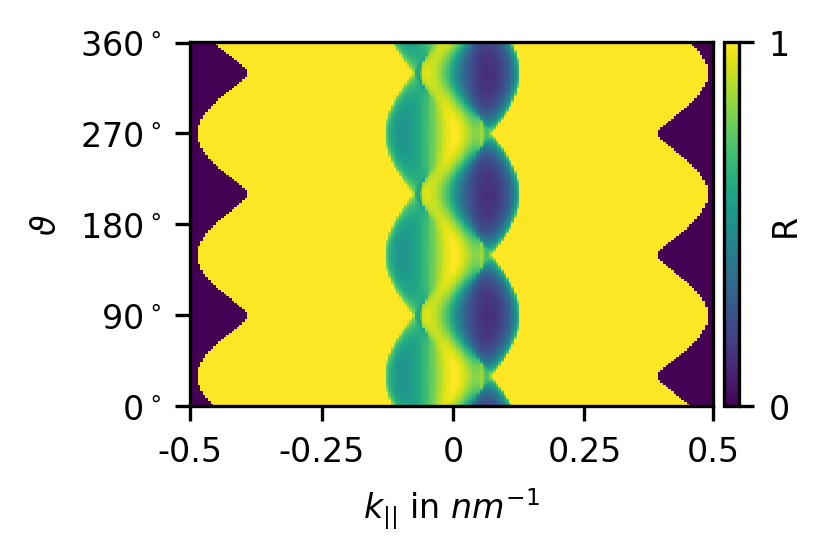}
    \caption{As Fig.~\ref{fig:map_0-5-10}(c), but for the other valley, $\xi=-1$. }
    \label{fig:app2}
\end{figure}

In Fig.~\ref{fig:app1} we show the phase space of Fig.~\ref{fig:model_system}(c) for the other valley. Figure \ref{fig:app2} indicates the reflectivity for the other valley. In short, the valleys are related by the symmetry transformation $k_\parallel \rightarrow - k_\parallel$ (mirroring the PSOS and reflectivity at the $k_\parallel=0$ line), whereas the Fermi lines of the two valleys are mirror symmetric with respect to the $k_y$ axis.

\clearpage
\newcommand{\newblock}{}
\bibliographystyle{unsrt}
\bibliography{BLGII}

\end{document}